\documentstyle[]{mn}
\newif\ifAMStwofonts

\def\etal{{\rm et al.}}

\def\simgt{\mathrel{\spose{\lower 3pt\hbox{$\sim$}}
        \raise 2.0pt\hbox{$>$}}}
\def\simlt{\mathrel{\spose{\lower 3pt\hbox{$\sim$}}
        \raise 2.0pt\hbox{$<$}}}

\ifoldfss
  \newcommand{\rmn}[1] {{\rm #1}}

  \ifCUPmtlplainloaded \else
    \NewTextAlphabet{textbfit} {cmbxti10} {}
    \NewTextAlphabet{textbfss} {cmssbx10} {}
    \NewMathAlphabet{mathbfit} {cmbxti10} {} 
    \NewMathAlphabet{mathbfss} {cmssbx10} {} 
  \fi
  \ifAMStwofonts
    \ifCUPmtlplainloaded \else
      \NewSymbolFont{upmath} {eurm10}
      \NewSymbolFont{AMSa} {msam10}
      \NewMathSymbol{\upi}     {0}{upmath}{19}
      \NewMathSymbol{\umu}     {0}{upmath}{16}
      \NewMathSymbol{\upartial}{0}{upmath}{40}
      \NewMathSymbol{\leqslant}{3}{AMSa}{36}
      \NewMathSymbol{\geqslant}{3}{AMSa}{3E}

    \fi
  \fi
\fi 

\ifnfssone
  \newmathalphabet{\mathit}
  \addtoversion{normal}{\mathit}{cmr}{m}{it}
  \addtoversion{bold}{\mathit}{cmr}{bx}{it}
  \newcommand{\rmn}[1] {\mathrm{#1}}

  \newmathalphabet{\mathbfit} 
  \addtoversion{normal}{\mathbfit}{cmr}{bx}{it}
  \addtoversion{bold}{\mathbfit}{cmr}{bx}{it}
  \newmathalphabet{\mathbfss} 
  \addtoversion{normal}{\mathbfss}{cmss}{bx}{n}
  \addtoversion{bold}{\mathbfss}{cmss}{bx}{n}
  \ifAMStwofonts
    \ifCUPmtlplainloaded \else
      \UseAMStwoboldmath
      \makeatletter
      \new@mathgroup\upmath@group
      \define@mathgroup\mv@normal\upmath@group{eur}{m}{n}
      \define@mathgroup\mv@bold\upmath@group{eur}{b}{n}
      \edef\UPM{\hexnumber\upmath@group}
      \new@mathgroup\amsa@group
      \define@mathgroup\mv@normal\amsa@group{msa}{m}{n}
      \define@mathgroup\mv@bold\amsa@group{msa}{m}{n}
      \edef\AMSa{\hexnumber\amsa@group}
      \makeatother
      \mathchardef\upi="0\UPM19
      \mathchardef\umu="0\UPM16
      \mathchardef\upartial="0\UPM40
      \mathchardef\leqslant="3\AMSa36
      \mathchardef\geqslant="3\AMSa3E
    \fi
  \fi
\fi 

\ifnfsstwo
  \newcommand{\rmn}[1] {\mathrm{#1}}

  \DeclareMathAlphabet{\mathbfit}{OT1}{cmr}{bx}{it}
  \SetMathAlphabet\mathbfit{bold}{OT1}{cmr}{bx}{it}
  \DeclareMathAlphabet{\mathbfss}{OT1}{cmss}{bx}{n}
  \SetMathAlphabet\mathbfss{bold}{OT1}{cmss}{bx}{n}
  \ifAMStwofonts
    \ifCUPmtlplainloaded \else
      \DeclareSymbolFont{UPM}{U}{eur}{m}{n}
      \SetSymbolFont{UPM}{bold}{U}{eur}{b}{n}
      \DeclareSymbolFont{AMSa}{U}{msa}{m}{n}
      \DeclareMathSymbol{\upi}{0}{UPM}{"19}
      \DeclareMathSymbol{\umu}{0}{UPM}{"16}
      \DeclareMathSymbol{\upartial}{0}{UPM}{"40}
      \DeclareMathSymbol{\leqslant}{3}{AMSa}{"36}
      \DeclareMathSymbol{\geqslant}{3}{AMSa}{"3E}
    \fi
  \fi
\fi 

\ifCUPmtlplainloaded \else
  \ifAMStwofonts \else 
    \def\upi{\pi}
    \def\umu{\mu}
    \def\upartial{\partial}
  \fi
\fi

\title[The rate of caustic crossing microlensing events for Q2237+0305]
  {The rate of caustic crossing microlensing events for Q2237+0305}
\author[J. S. B. Wyithe et al.]
  {J.~S.~B.~Wyithe,$^{1,2}$ 
  R.~L.~Webster$^1$, 
  E.~L.~Turner$^2$, \\
  $^1$ School of Physics, The University of Melbourne, Parkville, Vic, 3052, 
Australia\\
  $^2$ Princeton University Observatory, Peyton Hall, Princeton, NJ 08544, USA\\ 
 Email: swyithe@astro.Princeton.edu, rwebster@physics.unimelb.edu.au, elt@astro.Princeton.edu }
\date{Accepted. Received}
\pagerange{\pageref{firstpage}--\pageref{lastpage}}
\pubyear{1999}

\def\LaTeX{L\kern-.36em\raise.3ex\hbox{a}\kern-.15em
    T\kern-.1667em\lower.7ex\hbox{E}\kern-.125emX}

\begin{document}

\label{firstpage}

\maketitle

\begin{abstract}
Spectrophotometric observations of the gravitationally microlensed quasar\newline
 Q2237+0305 during a High Magnification Event (HME) is potentially a very powerful tool for probing the structure of the quasars accretion disc on scales of $\sim 10^{-8}$ arc seconds. In cases where the HME is produced by a single caustic (SHME), microlensing induced changes in the spectrum during the event may be used to directly infer the source intensity profile (eg. Grieger et al. 1988; Yonehara et al. 1988; Agol \& Krolik 1999a). Several groups are actively monitoring Q2237+0305 with this goal in mind.    
How often we can expect to observe a HME is dependent on the lens system parameters of galactic transverse velocity, mean microlens mass and the size of the magnified continuum source. We have previously used published microlensed light-curves to obtain expressions for the likely-hood of the values for these parameters (Wyithe, Webster \& Turner 1999b,c; Wyithe, Webster, Turner \& Mortlock 1999). Here we use this information to investigate the expected rate of SHMEs. We find the average rate of SHMEs as well as the number that we can expect to observe over periods of a decade and of a single observing season. We find that the average SHME rate summed over all images in Q2237+0305 is 1.5$\pm$0.6 - 6.2$\pm$1.3 events per decade. During the period following a caustic crossing we find that the event rate in the corresponding image is enhanced by $\sim50-100\%$, and therefore that the overall event rate may be higher during these periods. 
 From the distribution of events expected during a 6 month period we find that there is 1 chance in $\sim4-10$ of observing a SHME per observing season. The systematic dependence in these values arises from the different assumptions for smooth matter content, orientation of the galactic transverse velocity and the size of photometric error in the monitoring data. The results support continued monitoring of Q2237+0305 with the aim of obtaining detailed spectroscopic and photometric observations of a SHME.

\end{abstract}


\section{Introduction}

The object Q2237+0305 (Huchra et al. 1985) comprises a source quasar at a redshift of $z=1.695$ that is gravitationally lensed by a foreground galaxy with $z=0.0394$ producing 4 resolvable images with separations of $\sim 1''$.  Each of the 4 images are observed through the galactic bulge, which has a microlensing optical depth in stars that is of order unity (eg. Kent \& Falco 1988; Schneider et al. 1988; Schmidt, Webster \& Lewis 1998). In addition, the proximity of the lensing galaxy means that the effective transverse velocity may be high. The combination of these considerations make Q2237+0305 the ideal object from which to study microlensing. Indeed, Q2237+0305 is the only object in which cosmological microlensing has been confirmed (Irwin et.al 1989; Corrigan et.al 1991).

During a microlensed high magnification event (HME), regions of the source that are much smaller than the microlens Einstein radius are differentially amplified. Simulations by Wambsganss \& Paczynski (1991) showed that this produces an observed colour change that should be observable. Irwin et al. (1989) found evidence for such a colour change during the 1988 event in image A of Q2237+0305. Later Lewis et al. (1998) determined the ratios of emission line equivalent widths relative to one image. They show that the ratios vary between images at a single epoch, and that the ratio for a single image (image A) varies between two different epochs.
 This spectral change is interpreted as being due to the different spatial extents of the continuum and emission line regions being differentially amplified due to microlensing. 

While reverberation mapping provides a technique for determining the geometric structure of a QSO, typical sampling time-scales are several days which correspond to $2.5\times10^{15}cm/day$, comparable to the expected size of the accretion disc. Observation of a SHME provides a probe of the quasars central engine on scales at least an order of magnitude smaller than this. 
Two approaches have been taken to study microlensed quasar spectra. Firstly, several authors have calculated the effect of microlensing on the spectra and luminosity of model continuum regions (eg. Jaroszynski et al. (1992); Rauch \& Blandford (1992); Yonehara et al. (1998)). A different approach is to invert the light curve observed to directly obtain a surface brightness profile of the source (eg. Grieger et al. 1998, 1991; Agol \& Krolik 1999). Which ever approach is taken, good time resolution of observations ($\sim 1$ per day), and broad wavelength coverage are required for the duration of the SHME so that the continuum region can be probed with the finest possible resolution over a large range of spatial scales. It is therefore essential to frequently monitor the quasar in order to detect the onset of the HME as early as possible. Daily ground based monitoring of Q2237+0305 is underway at the Apache point observatory. In addition monitoring is implemented by the OGLE collaboration. The aim of the former program is to observe the onset of an HME so as to obtain spectra at different times during the event.

The rate of HMEs has been investigated by several authors (eg. Wambsganss, Paczynski \& Schneider (1990); Wambsganss, Paczynski \& Katz (1989); Witt, Kayser \& Refsdal 1993). However the expected rate of events has been poorly constrained due to the absence of information on several microlensing parameters.
 In particular the event rate is proportional to the effective galactic transverse velocity, and inversely proportional to the square root of the mean microlens mass. The size of the source will also effect the proportion of HMEs that are SHMEs. In this work we calculate HME and SHME rates at many assumed values for the above parameters, and convolve these rates with probabilities for the parameter values (Wyithe, Webster \& Turner 1999b,c; Wyithe, Webster, Turner \& Mortlock 1999).

This paper is presented in four parts Sections \ref{model} and \ref{rate} describe the microlensing model for Q2237+0305 and the calculation of the expected event rates respectively. Section \ref{results} describes the results obtained.

\section{The Microlensing model}
\label{model}

\begin{table}
\begin{center}
\caption{\label{params}Values of the total optical depth and the magnitude of the shear at the position of each of the 4 images of Q2237+0305. The quoted values are those of Schmidt, Webster \& Lewis (1998). $\kappa_{*}$ and $\kappa_{c}$ are the optical depths in stars and in smoothly distributed matter respectively.}
\begin{tabular}{|c|c|c|}
\hline
Image & $\kappa=\kappa_{*}+\kappa_{c}$   & $|\gamma|$  \\ \hline
  A   & 0.36                             &  0.40        \\
  B   & 0.36                             &  0.40        \\
  C   & 0.69                             &  0.71        \\
  D   & 0.59                             &  0.61        \\ \hline
\end{tabular}
\end{center}
\end{table}

 Throughout the paper, standard notation for gravitational lensing is used. The Einstein radius of a 1$M_{\odot} $ star in the source plane is denoted by $\eta_{0} $. The normalised shear is denoted by $\gamma$, and the convergence or optical depth by $\kappa$. $\kappa_{c}$ and $\kappa_{*}$ are the optical depth in smoothly distributed matter and stars respectively. To construct a microlensed light-curve we use the inversion technique of Lewis et al. (1993) and Witt (1993). For the microlensing models of Q2237+0305 presented in the current work we assume the macro-parameters for the lensing galaxy calculated by Schmidt, Webster \& Lewis (1998). These values are shown in table \ref{params}. 
 Where required a cosmology having $\Omega=1$ with $H_{o}=75\,km\,sec^{-1}$ is assumed. 
For our analysis we assume that microlensing is produced through the combination of a galactic transverse velocity with an isotropic velocity dispersion. Moreover, we assume that the magnitude of the dispersion is the same at the position of each of the four images. For the line-of-sight velocity dispersion of the stars in the galactic bulge we take the theoretical value of $\sigma_{*}\sim 165\,km\,sec^{-1}$ (Schmidt, Webster \& Lewis 1998). 

We describe the microlensing rate in terms of the effective transverse velocity which is defined as the transverse velocity that produces a microlensing rate from a static model equal to that of the observed light-curve (Wyithe, Webster \& Turner 1999b). The effective transverse velocity therefore describes the microlensing rate due to the combination of the effects of a galactic transverse velocity and proper motion of microlenses. When calculating HME rates we assume that the effective transverse velocity accurately describes not only the time averaged HME rate, but also the temporal clustering of HMEs and the distribution of orientations between the caustic and source trajectory.

We have previously obtained the following normalised probability distributions.

\noindent $i)$ $p_{s}(S_{\rmn s}|\langle m \rangle,v_{eff})$, the probability that the continuum source diameter is between $S_{\rmn s}$ and $S_{\rmn s} + {\rmn d}S_{\rmn s}$ given a mean microlens mass $\langle m \rangle$ and an effective galactic transverse velocity $v_{eff}$ (Wyithe, Webster, Turner \& Mortlock 2000).

\noindent $ii)$ $p_{v}(v_{eff}|\langle m \rangle)$ the probability that the effective galactic transverse velocity is between $v_{eff}$ and $v_{eff}+{\rmn d} v_{eff}$ given a mean microlens mass $\langle m \rangle$ (Wyithe, Webster \& Turner 1999b). 

\noindent $iii)$ $p_{m}(\langle m \rangle)$, the probability that the mean microlens mass is between $\langle m \rangle$ and $\langle m \rangle + {\rmn d} \langle m \rangle$ (Wyithe, Webster \& Turner 2000).

$p_{v}(v_{eff}|\langle m \rangle)$ and $p_{m}(\langle m \rangle)$ were computed using flat ($p(V_{tran})\propto dV_{tran}$), and logarithmic ($p(V_{tran})\propto \frac{dV_{tran}}{V_{tran}}$) assumptions for the Bayesian prior for galactic transverse velocity ($V_{tran}$). $p_{v}(v_{eff}|\langle m \rangle)$ was found to be insensitive to the prior assumed, however $p_{m}(\langle m \rangle)$ showed some dependence. For a point source, the rate of events calculated from $p_{v}(v_{eff}|\langle m \rangle)$ is independent of $p_{m}(\langle m \rangle)$, however  a slight dependence is introduced due to the finite size of the source in relation to the caustic network. In the remainder of this paper we use $p_{m}(\langle m \rangle)$ as calculated using the assumption of a logarithmic prior. We note that the assumption of the flat prior raises the computed event rates by a few percent.

 The functions $p_{s}(S_{\rmn s}|\langle m \rangle,v_{eff})$, $p_{v}(v_{eff}|\langle m \rangle)$, $p_{m}(\langle m \rangle)$ and the event rates presented in this paper were computed for the following assumptions of smooth matter density, photometric error, and direction of the galactic transverse velocity.
Two models are considered for the distribution of microlenses, one with no continuously distributed matter, and one where smooth matter contributes 50\% of the surface mass density. 
 Two orientations were chosen for the transverse velocity with respect to the galaxy, with the source trajectory being parallel to the A$-$B and C$-$D axes.
 The two orientations bracket the range of possibilities, and because the images are positioned approximately orthogonally with respect to the galactic centre correspond to shear values of $\gamma_{A},\gamma_{B}<0,\gamma_{C},\gamma_{D}>0$ and $\gamma_{A},\gamma_{B}<0,\gamma_{C},\gamma_{D}>0$ respectively.
 The simulations used two different estimates of the error in the photometric magnitudes. In the first case a small error was assumed (SE). For images A and B, $\sigma_{SE}$=0.01 mag, and for images C and D $\sigma_{SE}$=0.02 mag. In the second case, a larger error was assumed (LE). For images A and B, $\sigma_{LE}$=0.02 mag and for images C and D $\sigma_{LE}$=0.04 mag. The observational error in Irwin et al. (1989) was 0.02 mag. 

 Both the microlensing rate due to a transverse velocity (eg. Witt, Kaiser \& Refsdal 1993), as well as the corresponding rate due to proper motions (Wyithe, Webster \& Turner 1999a) are not functions of the details of the microlens mass distribution, but rather are only dependent on the mean microlens mass. We therefore limit our attention to models in which all the point masses have the same mass since the results obtained will be applicable to other models with different forms for the mass function.
The microlensing models presented in this work have been discussed in detail in Wyithe, Webster \& Turner (2000).

We obtain caustic positions and orientations as follows using the contouring algorithm (Witt 1993; Lewis et al. 1993), which we implement according to the description of Lewis et al. (1993). The image curves are followed in discrete steps in one direction. During the contouring algorithm the situation of the image line crossing a critical curve results in the image magnification changing sign due to a parity flip between the critical images on either side of the critical curve. The location of the critical curve can then be found through a search for the image position that produces a  lens mapping  Jacobian determinant of zero. This value for the critical image position is used to calculate the corresponding caustic position. The gradient of the caustic tangent is found by first computing the location of a point a short distance along the critical curve in both directions. The corresponding caustic points are then used to compute the gradient of the caustic tangent from a three-point derivative.

\begin{table*}
\begin{center}
\caption{\label{table} The means and variances of the probability distribution functions for HME and SHME rates (events/decade).}
\begin{tabular}{|c|c|c|c|c|c|c|c|c|}

\hline
                                          &   \multicolumn{4}{c}{$\gamma_{A,B}>0,\gamma_{C,D}<0$} &  \multicolumn{4}{c}{$\gamma_{A,B}<0,\gamma_{C,D}>0$} \\
         Error ($\Delta m=\pm0.02$)       & \multicolumn{2}{c}{0\%$\kappa_{c}$}  & \multicolumn{2}{c}{50\%$\kappa_{c}$} & \multicolumn{2}{c}{0\%$\kappa_{c}$}  & \multicolumn{2}{c}{50\%$\kappa_{c}$} \\
                                          & HME rate  & SHME rate & HME rate  & SHME rate & HME rate  & SHME rate & HME rate  & SHME rate \\\hline\hline 

  $2\sigma_{A,B},1\sigma_{C,D}$           &  7.3$\pm$1.5 &  6.2$\pm$1.3    &   5.6$\pm$1.3   & 4.8$\pm$1.3   &  4.3$\pm$1.2  &  3.5$\pm$1.0    &   2.8$\pm$0.8   &    2.3$\pm$0.7  \\
  $1\sigma_{A,B},\frac{1}{2}\sigma_{C,D}$ &  4.8$\pm$1.5  & 4.1$\pm$.4      &   3.7$\pm$1.3   & 3.2$\pm$1.1  &  2.8$\pm$1.0  &  2.3$\pm$0.8    & 1.8$\pm$0.7     &    1.5$\pm$0.6  \\\hline

\end{tabular}
\end{center}
\end{table*}

\begin{table*}
\begin{center}
\caption{\label{table_2} The means of the histograms $h_N(N)$ for SHME rates (events/decade) in the cases of events during randomly chosen 10 year periods (Uncorr. rate) and 10 year periods imediately following a caustic crossing (Corr. rate).}
\begin{tabular}{|c|c|c|c|c|c|c|c|c|}

\hline
                                          &   \multicolumn{4}{c}{$\gamma_{A,B}>0,\gamma_{C,D}<0$} &  \multicolumn{4}{c}{$\gamma_{A,B}<0,\gamma_{C,D}>0$} \\
         Error ($\Delta m=\pm0.02$)       & \multicolumn{2}{c}{0\%$\kappa_{c}$}  & \multicolumn{2}{c}{50\%$\kappa_{c}$} & \multicolumn{2}{c}{0\%$\kappa_{c}$}  & \multicolumn{2}{c}{50\%$\kappa_{c}$} \\
                                          & Uncorr rate  & Corr rate & Uncorr rate & Corr rate & Uncorr rate  & Corr rate & Uncorr rate & Corr rate \\\hline\hline 
Image A/B                                 &              &                 &                 &               &               &                &                  &                \\\hline                  

  $2\sigma_{A,B},1\sigma_{C,D}$           &    1.5       &    2.3          &  1.2            & 1.9           &   1.0         & 1.7             &   0.6           &   1.2           \\
  $1\sigma_{A,B},\frac{1}{2}\sigma_{C,D}$ &   1.0         &  1.9            &  0.9            &  1.6         &   0.7         &1.5              &  0.5            &   1.1           \\\hline

Image C                                 &              &                 &                 &               &               &                &                  &                \\\hline                  
  $2\sigma_{A,B},1\sigma_{C,D}$           & 1.4          &   2.2           &      0.9        &  1.5          &  0.9          &  1.5            &  0.5            &   1.1           \\
  $1\sigma_{A,B},\frac{1}{2}\sigma_{C,D}$ &  1.0          &   1.9           &   0.7           &   1.4        &  0.6          &  1.3            &   0.4           &   1.1           \\\hline

Image D                                 &              &                 &                 &               &               &                &                  &                \\\hline                  
  $2\sigma_{A,B},1\sigma_{C,D}$           &   2.3        &   3.3           &  2.0            &  2.7          &  1.3          &    2.0          &    1.0          &    1.6          \\
  $1\sigma_{A,B},\frac{1}{2}\sigma_{C,D}$ &   1.7         &  2.6            &  1.4            &  2.2         &   0.9         & 1.7             &     0.7         &     1.5         \\\hline

\end{tabular}
\end{center}
\end{table*}

\section{Event rate}
\label{rate}
We find the rate of caustic crossings as a function of mean microlens mass, source size and transverse velocity. We define two classes of HMEs for this purpose: 1) those that involve a single caustic and so can be deconvolved to yield information on the source intensity profile etc. of the source (SHME), and 2) those where the source is directly effected by two or more caustics at once (MHME). The following algorithm is used to automatically classify a model HME into one of these classes. We define a caustic to be involved in an event if the closest edge of the source is within one source diameter of the caustic on the + side (source has associated critical images), or in contact with the caustic on the - side (source has no associated critical images). It is computationally prohibitive to compute full light curves for extended sources of many sizes over a statistically representitive number of starfields. Therefore, we make the following approximation for the small sources ($\ll 1\eta_{o}$) that we are considering here. We assume that the caustic is locally straight over a length that is larger than the source. If the caustic makes an angle with the source track of $\theta$, then a circular source comes into contact with it at a distance of $L=\frac{S}{2\sin{\theta}}$ along the source line from the source centre. This length can be calculated for the nearest caustic along the source line in both directions ($L_{left},L_{right}$). The source is only effected by one caustic at a time if the distance between these caustics along the source line is less than $L_{left}+L_{right}$ for $+-$, $2S+L_{left}+L_{right}$ for $-+$ and $S+L_{left}+L_{right}$ for $++$ or $--$ caustic pairs. If this condition is fulfilled for a source on both sides of a caustic then the source is said to undergo a SHME when crossing that caustic.

The average rate of SHMEs for a given source size, microlens mass and effective galactic transverse velocity is 
\begin{equation}
\nonumber
\bar{R}\left(S,\langle m\rangle\right)=\frac{v_{eff}\left(\langle m\rangle\right)}{\sqrt{\langle m\rangle}}\bar{R}_{o}\left(S,\langle m\rangle\right)
\end{equation}
where $\bar{R}_{o}\left(S,\langle m\rangle\right)$ is the rate for $v_{eff}\left(\langle m\rangle\right)=1$.
The probability of $\bar{R}\left(S,\langle m\rangle\right)$ lying between $\bar{R}$ and $\bar{R}+\Delta \bar{R}$ is therefore
\begin{eqnarray}
\nonumber
&&\hspace{-5mm}p_{\bar{R}}\left(\bar{R}|S,\langle m\rangle\right)=\\
&&\hspace{5mm}p_{v}\left(v_{eff}=\frac{\sqrt{\langle m\rangle}\bar{R}}{\bar{R}_{o}\left(S,\langle m\rangle\right)}|\langle m\rangle\right)\frac{\sqrt{\langle m\rangle}}{\bar{R}_{o}\left(S,\langle m\rangle\right)}
\end{eqnarray}
From this we can evaluate the probability for the SHME rate given our previous calculations for the probability of microlens mass and source size:
\begin{eqnarray}
\nonumber
&&\hspace{-5mm}p_{\bar{R}}(\bar{R})=\\
&&\hspace{-3mm}\int dm\int dS\,\, p_{s}\left(S|\langle m\rangle,v_{eff}\right)p_{m}\left(\langle m\rangle\right)p_{\bar{R}}\left(\bar{R}|S,\langle m\rangle\right)
\end{eqnarray}
where
\begin{equation}
v_{eff}=\frac{\sqrt{\langle m\rangle}\bar{R}}{\bar{R}_{o}\left(S,\langle m\rangle\right)}.
\end{equation}
The caustic network is clustered on temporal scales of decades. Therefore, the average event rate may not be observable for as long as a century (Wambsganss, Paczynski \& Schneider 1990). For this reason the distribution of the number of SHMEs expected during a finite period is of more utility.
 We calculate the histogram for the number of SHMEs per period for a collection of assumed effective galactic transverse velocities, source sizes and mean microlens masses:
\begin{equation} 
h_{N}\left(N|S,\langle m\rangle,v_{eff}\right).
\end{equation}
The probability for finding a number of SHMEs in a period can then be obtained:
\begin{eqnarray}
\nonumber
&&\hspace{-7mm}h_{N}\left(N\right)=\\
\nonumber&&\hspace{-3mm}\int dm\int dS\int dv_{eff}\,\,\left( p_{s}\left(S|\langle m\rangle,v_{eff}\right)p_{m}\left(\langle m\rangle\right)\right.\\
&&\hspace{20mm}\times\left.p_{v}\left(v_{eff}|\langle m\rangle\right)h_{N}\left(N|S,\langle m\rangle,v_{eff}\right)\right)
\end{eqnarray}

$p_{\bar{R_{i}}}(\bar{R_{i}})$ and $h_{N_{i}}\left(N_{i}\right)$ are evaluated for each of the four images $(i=1-4)$. The total rate summed over all images is then evaluated as
\begin{equation}
p_{\bar{R}_{tot}}\left(\bar{R}_{tot}\right)=\int \Pi_{i=1}^{4} d\bar{R}_{i} p_{\bar{R}_{i}}\left(\bar{R}_{i}\right)\delta\left(\bar{R}_{tot}-\left(\Sigma_{i=1}^{4}\bar{R}_{i}\right)\right).
\end{equation}
Similarly, the probability for finding a number of SHMEs in a finite period summed over all images is  
\begin{equation}
h_{N_{tot}}\left(N_{tot}\right)=\sum \Pi_{i=1}^{4}h_{N_{i}}\left(N_{i}\right)\delta\left({N_{tot},\Sigma_{i=1}^{4}N_{i}}\right). 
\end{equation}

Monitoring data (Lens Monitoring Project, Apache Point Observatory; OGLE collaboration (see http://www.astro.princeton.edu/ogle/ogle2/huchra.html)) shows evidence for recent high magnification events in two images. Clustering of caustics will tend to produce a higher than average rate of caustic crossings (in a single image) in the period following a HME. In addition to calculating $h_{N}\left(N\right)$ for randomly chosen 10 year periods, we therefore also calculate $h_{N}\left(N\right)$ for a sample 10 year periods that immediately follow a caustic crossing.

\begin{figure*}
\vspace*{100mm}
\includegraphics{fig1.epsi}
\caption{\label{0smooth_images}The functions representing the probability for event rates in Q2237+0305. Left: the time averaged rate, with expectation values, Centre: events expected per decade and Right: events expected per 6 month observing season for images A/B, C and D. The light and dark lines correspond to $\gamma<0$ and $\gamma>0$ respectively, the thick and thin lines represent the resulting functions when the photometric error was assumed to be $\sigma_{SE}$=0.01 mag in images A/B and $\sigma_{SE}$=0.02 in images C/D, and $\sigma_{LE}$=0.02 mag in images A/B and $\sigma_{LE}$=0.04 in images C/D. The lens model contained no smoothly distributed matter.}
\vspace*{40mm}
\includegraphics{fig2.epsi}
\caption{\label{0smooth_tot}The functions representing the probability for combined image event rates in Q2237+0305. Left: the time averaged rate, with expectation values, Centre: events expected per decade and Right: events expected per 6 month observing season. The light and dark lines refer to assumed directions for transverse motion that are directed along the A-B axis and C-D axes respectively. The thick and thin lines represent the resulting functions when the photometric error was assumed to be $\sigma_{SE}$=0.01 mag in images A/B and $\sigma_{SE}$=0.02 in images C/D, and $\sigma_{LE}$=0.02 mag in images A/B and $\sigma_{LE}$=0.04 in images C/D. The lens model contained no smoothly distributed matter.}
\end{figure*}

\begin{figure*}
\vspace*{100mm}
\includegraphics{fig3.epsi}
\caption{\label{50smooth_images}The functions representing the probability for event rates in Q2237+0305. Left: the time averaged rate, with expectation values, Centre: events expected per decade and Right: events expected per 6 month observing season for images A/B, C and D. The light and dark lines correspond to $\gamma<0$ and $\gamma>0$ respectively, the thick and thin lines represent the resulting functions when the photometric error was assumed to be $\sigma_{SE}$=0.01 mag in images A/B and $\sigma_{SE}$=0.02 in images C/D, and $\sigma_{LE}$=0.02 mag in images A/B and $\sigma_{LE}$=0.04 in images C/D. The lens model contained 50\% smoothly distributed matter.}
\vspace*{40mm}
\includegraphics{fig4.epsi}
\caption{\label{50smooth_tot}The functions representing the probability for combined image event rates in Q2237+0305. Left: the time averaged rate, with expectation values, Centre: events expected per decade and Right: events expected per 6 month observing season. The light and dark lines refer to assumed directions for transverse motion that are directed along the A-B axis and C-D axes respectively. The thick and thin lines represent the resulting functions when the photometric error was assumed to be $\sigma_{SE}$=0.01 mag in images A/B and $\sigma_{SE}$=0.02 in images C/D, and $\sigma_{LE}$=0.02 mag in images A/B and $\sigma_{LE}$=0.04 in images C/D. The lens model contained 50\% smoothly distributed matter.}
\end{figure*}

\begin{figure*}
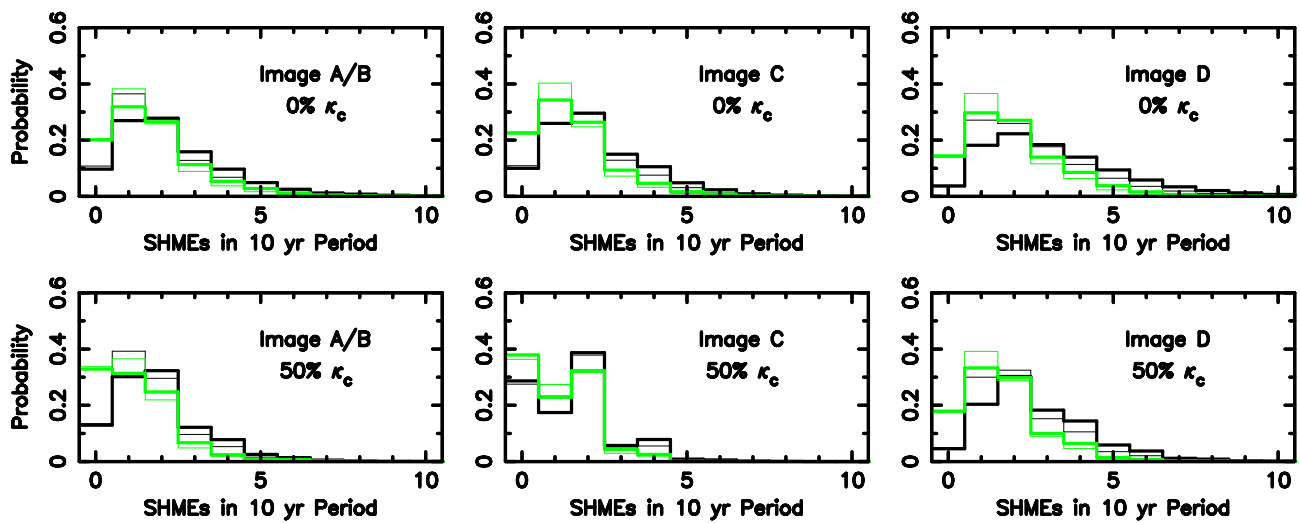

\vspace{70mm}
\includegraphics{fig5a.epsi}
\includegraphics{fig5b.epsi}
\caption{\label{correlated}The functions representing the probability for events expected in the decade following the observation of an HME.
The light and dark lines refer to assumed directions for transverse motion that are directed along the A-B axis and C-D axes respectively. The thick and thin lines represent the resulting functions when the photometric error was assumed to be $\sigma_{SE}$=0.01 mag in images A/B and $\sigma_{SE}$=0.02 in images C/D, and $\sigma_{LE}$=0.02 mag in images A/B and $\sigma_{LE}$=0.04 in images C/D. }
\end{figure*}

\section{results}
\label{results}

Table \ref{table} contains the expectation values and variances for the total HME and SHME rate. The variances in individual distributions are comparable to the systematic uncertainty in the expectation value introduced by the assumptions of trajectory direction, photometric error, and smooth matter content. The rates for HMEs and SHMEs are similar in all cases, with SHMEs being slightly less common. The small difference between these values is a result of the low probability for a source size that is $>0.1\sqrt{\langle m\rangle}\eta_{o}$. The values range between 1.5$\pm$0.6 and 6.2$\pm$1.3 events per decade. These rates are consistent with those calculated by Witt, Kayser \& Refsdal (1993) who assumed a source plane transverse velocity of 6000$\,km\,sec^{-1}$ and a mean microlens mass of $\sim 0.35M_{\odot}$. The rates are lower where the trajectory is aligned with the A-B axis. This effect is larger than the lowering of the event rate that is produced by the introduction of smooth matter. In addition, the predicted event rate is dependent on the photometric error assumed. This dependence comes from the systematic uncertainty that the assumption introduces into the determination of the effective transverse velocity.

 The microlensing parameter distributions were calculated from an ensemble of 5000 simulations of the observed data for each image, and for each assumption of trajectory direction, photometric error, and smooth matter content. The 5000 simulations were spread over 1000$\eta_{o}$ of simulated light curves. If the error in the average rate is assumed to be due to Poison noise then it is less than $\sim 1\%$. The errors in the modelling statistics are therefore not important in determining the probability distributions for the average HME rate.    

Figures \ref{0smooth_images}, \ref{0smooth_tot}, \ref{50smooth_images} and \ref{50smooth_tot} display the functions $p_{\bar{R}}(\bar{R})$, $p_{\bar{R}_{tot}}(\bar{R}_{tot})$, $h_{N}\left(N\right)$ and $h_{N_{tot}}\left(N_{tot}\right)$ obtained for the microlensing models considered. Both the rates for the individual images as well as the combined rates are shown. $h_{N}\left(N\right)$ and $h_{N_{tot}}\left(N_{tot}\right)$ have been calculated for periods of 1 decade, corresponding approximately to the total Q2237+0305 monitoring period, as well as a 6 month period which corresponds to a single observing season. 

We find that the event rates should be highest in image D, corresponding to the higher expected magnification. This is in contrast to the observed behaviour of image D which has been relatively faint for the period of monitoring, suggesting that the source  may currently reside in a sparse region of the corresponding caustic network. It is important to note however that the non-observation of a HME in any image over the time that Q2237+0305 has been monitored should not be surprising, since the modes of all individual image histograms are zero. This offers an alternative to the explanation given by Witt \& Mao (1994) that the lack of variability in images C and D (which persisted following the publication of that paper) is the result of an alignment between the transverse velocity and the caustic clustering. 

The probability distributions of microlensing model parameters that have been used to calculate the event rates have been obtained from the observed light curves. However these statistics are derived primarily from the rate of low level microlensing variation which is related to but does not necessarily correspond directly with the rate of caustic crossings. It is therefore interesting to compare the number of HMEs in the observed light curves to the calculated distribution. There is circumstantial evidence for approximately 5 HMEs in existing light-curves for Q2237+0305 (Irwin et al. 1989; Corrigan et al. 1991; $\O$stensen et al. 1995; OGLE Web page). This number is larger than, but consistent with the mode of most of the distributions. However, it may be improbably high if the combination of a trajectory along the A-B axis and 50\% smooth matter content is assumed (for both assumptions of photometric error).

There is 1 chance in $\sim 4-10$ (20 for cases where a trajectory along the A-B axis and 50\% smooth matter content are assumed) that a SHME will be observed during a given observing season. However the observation of two SHMEs in one observing season is very unlikely ($<1/25$), indicating that the temporal scale of a diamond shaped caustic is typically larger than 6 months. Witt, Kayser \& Refsdal (1993) found that M-shaped double events have an average separation of 0.9-1.3 years (they assumed a source plane transverse velocity of 6000$\,km\,sec^{-1}$ and a mean microlens mass of $\sim 0.35$). This is demonstrated by the probable double peaked event in image A (Irwin et al. 1989; Corrigan et al. 1991; Racine 1992) which has a duration of $\sim 500$ days.

Figure \ref{correlated} shows histograms of the number of SHMEs expected in the 10 years immediately following the observation of an HME. The figure shows a large excess in the rate of events during such periods. Indeed, the most likely number of events is non-zero. The means of these histograms, as well as the corresponding means computed for the average rate are shown in table~\ref{table_2}. The values demonstrate a $\sim50-100\%$ increase in SHME rate in the decade following an HME over the average value.

Two images in Q2237+0305 have undergone a significant change in brightness between the end of the $\O$stensen et al. (1996) data and the beginning of the 1998 observing season. The expected rate of SHMEs for the next 10 years may therefore be significantly higher than the values presented in table \ref{table}.

\section{Conclusion}
We have combined calculations of the time averaged rates of SHMEs over a phase space of assumed values for effective galactic transverse velocity, mean microlens mass and continuum source size with previously obtained probabilities for their values. We find that the total SHME rate in Q2237+0305 is 1.5$\pm$0.6 - 6.2$\pm$1.3 events per decade depending on the assumptions made for the direction of the galactic transverse velocity, the smooth matter content and photometric error. We find that the probable number of  observed HMEs is consistent with calculated distributions of events per 10 year period. During periods following a caustic crossing in an image, we find that the event rate in that image is between $\sim$50 and 100\% higher than the average value.
In addition, we have calculated the distribution of SHMEs that should be observed per 6 month period and find that there is one chance in $\sim4-10$ that a SHME will be observed. The 6 month period corresponds approximately to one observing season. These results therefore add further support to the case for continued monitoring of Q2237+0305 with the aim of making spectroscopic and photometric observations of Q2237+0305 during a SHME.

\label{lastpage}


\begin{thebibliography}{}

\bibitem[\protect\citename{Agol \& Krolik}1999]{AG99}
Agol, E., Krolik, J., 1999, Ap.  J., in press (astro-ph/9905111)

\bibitem[\protect\citename{Corrigan \etal }1991]{CO91}
Corrigan et.al, 1991, Astron. J., 102, 34
 
\bibitem[\protect\citename{Grieger et al. }1991]{GR91}
Grieger, B., Kayser, R., Schramm, T., 1991, Astron. Astrophys., 252, 508

\bibitem[\protect\citename{Grieger et al. }1988]{GR88}
Grieger, B., Kayser, R., Refsdal, S., 1988, Astron. Astrophys., 194, 54

\bibitem[\protect\citename{Huchra et al. }1985]{HU85}
Huchra, J., Gorenstein, M., Horine, E., Kent, S., Perley, R., Shapiro, I. I., Smith, G., 1985, Astron. J., 90, 691

\bibitem[\protect\citename{Irwin \etal  }1989]{IR89}
Irwin, M. J., Webster, R. L., Hewitt, P. C., Corrigan, R. T., Jedrzejewski, R. I., 1989, Astron. J., 98, 1989
 
\bibitem[\protect\citename{Jaroszynski, Wambsganss \& Paczynski }1992]{JA92}
Jaroszynski, M., Wambsganss, J., Paczynski, B., 1992, Ap. J., 396, L65 

\bibitem[\protect\citename{Kent \& Falco }1988]{KE88}
Kent, S. M., Falco, E. E., 1988, Astron. J., 96, 1570

\bibitem[\protect\citename{Lewis, Irwin, Hewett \& Foltz }1998]{LW98}
Lewis, G. F., Irwin, M. J., Hewett, P. C., Foltz, C. B., 1998, MNRAS, accepted

\bibitem[\protect\citename{Lewis et al.   }1993]{LE93}
Lewis, G. F., Miralda-Escude, J., Richardson, D. C., Wambsganss, J., 1993, MNRAS, 261, 647

\bibitem[\protect\citename{$\O$stensen \etal}1996]{OS96}
$\O$stensen, R. \etal 1996, Astron. Astrophys., 309, 59

\bibitem[\protect\citename{Racine } 1992]{RC92}
Racine, 1992, Ap. J., 395, L65

\bibitem[\protect\citename{Rauch \& Blandford } 1991]{RA92}
Rauch, K. P., Blandford, R. D., 1991, Ap. J. 1991, 381, L39

\bibitem[\protect\citename{Schmidt, Webster \& Lewis }1998]{SC98}
Schmidt, R. W., Webster, R. L., Lewis, G. F. 1998, MNRAS, 295, 488

\bibitem[\protect\citename{Schneider et al. }1988]{SC88}
Schneider, D. P., Turner, E. L., Gunn, J. E., Hewitt, J. N., Schmidt, M., Lawrence, C. R., 1988, Astron. J., 95, 1619 

\bibitem[\protect\citename{Wambsganss \& Paczynski}1991]{WA91}
Wambsganss, J., Paczynski, B., 1991, Astron. J., 102, 864 
 
\bibitem[\protect\citename{Wambsganss, Paczynski \& Katz }1989]{WA89}
Wambsganss, J., Paczynski, B., Katz, N., 1989, Ap. J., 352, 407 
 
\bibitem[\protect\citename{Wambsganss, Paczynski \& Schneider }1990]{WA90b}
Wambsganss, J., Paczynski, B., Schneider, P., 1990, Ap. J., 358, L33

\bibitem[\protect\citename{Witt }1993]{WI93}
Witt, H. J., 1993, Ap. J., 403, 530

\bibitem[\protect\citename{Witt, Kayser \& Refsdal }1993]{WI93}
Witt, H. J., Kayser, R., \& Refsdal, S. 1993, Astron. Astrophys.,268, 501

\bibitem[\protect\citename{Witt \& Mao }1994]{WI94}
Witt, H. J., Mao, S., 1994, Ap. J., 429, 66

\bibitem[\protect\citename{Wyithe \& Webster}1999]{WY99}
Wyithe, J. S. B, Webster, R. L., 1999, MNRAS, 306, 261

\bibitem[\protect\citename{Wyithe, Webster \& Turner }1999a]{WY98a}
Wyithe, J. S. B, Webster, R. L., Turner, E. L., 1999a, MNRAS, 312, 843

\bibitem[\protect\citename{Wyithe, Webster \& Turner }1999b]{WY98b}
Wyithe, J. S. B, Webster, R. L., Turner, E. L., 1999b, MNRAS, 309, 261

\bibitem[\protect\citename{Wyithe, Webster \& Turner }1999c]{WY98c}
Wyithe, J. S. B, Webster, R. L., Turner, E. L., 2000, MNRAS 315, 51

\bibitem[\protect\citename{Wyithe, Webster, Turner \& Mortlock}1999c]{WY98c}
Wyithe, J. S. B, Webster, R. L., Turner, E. L., D. J. Mortlock, 2000, MNRAS, 315, 62

\bibitem[\protect\citename{Yonehara et al.}1998]{YO98}
Yonehara, A., et al., 1998, Ap. J., 501, L41

\end{thebibliography}
\end{document}